\documentclass[conference]{IEEEtran}
\makeatletter
\renewcommand{\@IEEEsectpunct}{\ \,}

\usepackage{stfloats}
\usepackage{geometry}
\usepackage{amsfonts}
\usepackage{amssymb}
\usepackage{amsthm}
\usepackage{cite}
\usepackage[cmex10]{amsmath}
\usepackage{float}
\usepackage{color}
\usepackage{stfloats,fancyhdr}
\usepackage{amsmath,bm}
\usepackage{algorithm}
\usepackage{algorithmic}
\usepackage{multirow}
\usepackage{changepage}
\usepackage[normalem]{ulem}

\IEEEoverridecommandlockouts
\ifCLASSINFOpdf
\usepackage[pdftex]{graphicx}
\DeclareGraphicsExtensions{.pdf,.jpeg,.png}
\else
\usepackage[dvips]{graphicx}
\DeclareGraphicsExtensions{.eps}
\fi

\usepackage{subfigure}
\usepackage{fancybox,dashbox}

\begin{document}
	
	\title{Full-Duplex Cooperative Cognitive Radio Networks with Wireless Energy Harvesting}
	\author{\IEEEauthorblockN{Rui Zhang, He Chen, Phee Lep Yeoh, Yonghui Li, and Branka Vucetic}
		\IEEEauthorblockA{School of Electrical and Information Engineering, University of Sydney, Australia\\
			Email: \{rui.zhang1, he.chen, phee.yeoh, yonghui.li, branka.vucetic \}@sydney.edu.au}
	}
	
	\maketitle
	\begin{abstract}
		This paper proposes and analyzes a new full-duplex (FD) cooperative cognitive radio network with wireless energy harvesting (EH). We consider that the secondary receiver is equipped with a FD radio and acts as a FD hybrid access point (HAP), which aims to collect information from its associated EH secondary transmitter (ST) and relay the signals. The ST is assumed to be equipped with an EH unit and a rechargeable battery such that it can harvest and accumulate energy from radio frequency (RF) signals transmitted by the primary transmitter (PT) and the HAP. We develop a novel cooperative spectrum sharing (CSS) protocol for the considered system. In the proposed protocol, thanks to its FD capability, the HAP can receive the PT's signals and transmit energy-bearing signals to charge the ST simultaneously, or forward the PT's signals and receive the ST's signals at the same time. We derive analytical expressions for the achievable throughput of both primary and secondary links by characterizing the dynamic charging/discharging behaviors of the ST battery as a finite-state Markov chain. We present numerical results to validate our theoretical analysis and demonstrate the merits of the proposed protocol over its non-cooperative counterpart.
	\end{abstract}
	
	\section{Introduction}
	\let\thefootnote\relax\footnote{This work was supported in part by ARC grants FL160100032, DP150104019, FT120100487 and DE140100420.}
	The performance of wireless communication systems is severely restricted by fundamental constraints on the energy and bandwidth of wireless transceivers. Motivated by this issue, incorporating wireless energy harvesting (EH) into cooperative cognitive radio networks with spectrum sharing has emerged as a promising solution to boost both the network lifetime and spectrum efficiency of future communication systems \cite{chen2016cooperative,6449254}. In this context, both the primary and secondary users can harvest energy from radio-frequency (RF) signals and use the harvested energy to perform information transmission/relaying. Moreover, the unlicensed (secondary) users can help relay the licensed (primary) user's information in exchange for permission to transmit their own information using a proportion of the same spectrum. The design and analysis of EH-based cooperative spectrum sharing (CSS) networks have attracted considerable attention in the open literature recently, see e.g., \cite{7169619,7342976,6847192,7434602} and references therein.
	
	Reference \cite{7169619} applied tools from stochastic geometry to study a large-scale EH-based CSS network with EH primary transmitters (PTs) that harvest energy from their associated hybrid access points (HAPs) and nearby secondary transmitters (STs). The secondary system performs energy cooperation with the primary HAPs to charge their EH PTs in exchange for a fraction of bandwidth. It was shown in \cite{7169619} that introducing cooperation between the primary and secondary systems can help improve the average throughput of both systems. Moreover, an alternative EH-based CSS network model was investigated in \cite{7342976,6847192,7434602}, wherein the ST was considered as an EH node that fully relies on the harvested energy to relay the PT's data and transmit its own data. In \cite{7342976}, the ST harvests energy and receives information from the PT at the same time using a power splitting technique, and broadcasts a superposition of the PT and its own information. A time switching-based CSS protocol was examined in \cite{6847192}, where the ST performs EH, information relaying, and its own information transmission in different time durations of each transmission block. In \cite{7434602}, the energy accumulation process at the EH ST was characterized, in which the working modes of the ST depend on its residual energy and the decoding status of the PT's data via the primary direct link.
	
	The network frameworks proposed in \cite{7342976,6847192,7434602} are promising since they enable the secondary users to harvest both energy and spectrum from the primary system, which could find considerable applications in existing and upcoming networks with low-energy and low-cost devices. However, the ST in these frameworks is burdened with not only the information relaying for PT(s) but also its own information transmission, which may unduly constrain the network performance. This is because it normally harvests and accumulates a limited amount of energy as a result of the significant propagation loss of RF signals. Motivated by this, in this paper we propose a new EH-based CSS network framework, where the relaying operation is assigned to the secondary receiver (SR) and the ST uses the harvested energy to transmit its own information only. To further boost the spectral efficiency, the full-duplex (FD) technique was recently shown in \cite{6493535} to allow wireless devices to transmit and receive on the same frequency band at the same time. Inspired by this, we consider that the SR is a FD HAP such that it can receive the PT's signals and transmit energy-bearing RF signals to charge the ST simultaneously. When the accumulated energy at the ST satisfies a predefined threshold, the ST can transmit information to the HAP, which can also relay the PT's signals at the same time thanks to its FD~capability.
	
	The main contributions of this paper are summarized as follows: (1) We propose a new CSS protocol for a FD cooperative cognitive radio network with wireless EH, in which the ST is an EH node that harvests energy from both the PT and the secondary HAP before transmitting its own information to the HAP; (2) We characterize the energy accumulation process of the ST battery, in which the harvested energy is stored over a certain number of transmission blocks, and is utilized for information transmission to the HAP only when a predefined energy threshold is reached. We model the dynamic behaviors of the ST battery as a finite state Markov chain and attain its steady state distribution; (3) We derive new analytical expressions for the achievable throughput of both primary and secondary systems over Nakagami-$m$ fading channels with integer $m$, which provides a good approximation to the line-of-sight (LoS) Rician channels that are typical in wireless EH scenarios \cite{1093888}; (4) Numerical results are performed to verify all theoretical analysis and illustrate the effects of key parameters on the derived system~performance.
	
	\section{System Model}
\begin{figure}[h]
		\centering
		\includegraphics[width=0.45\textwidth]{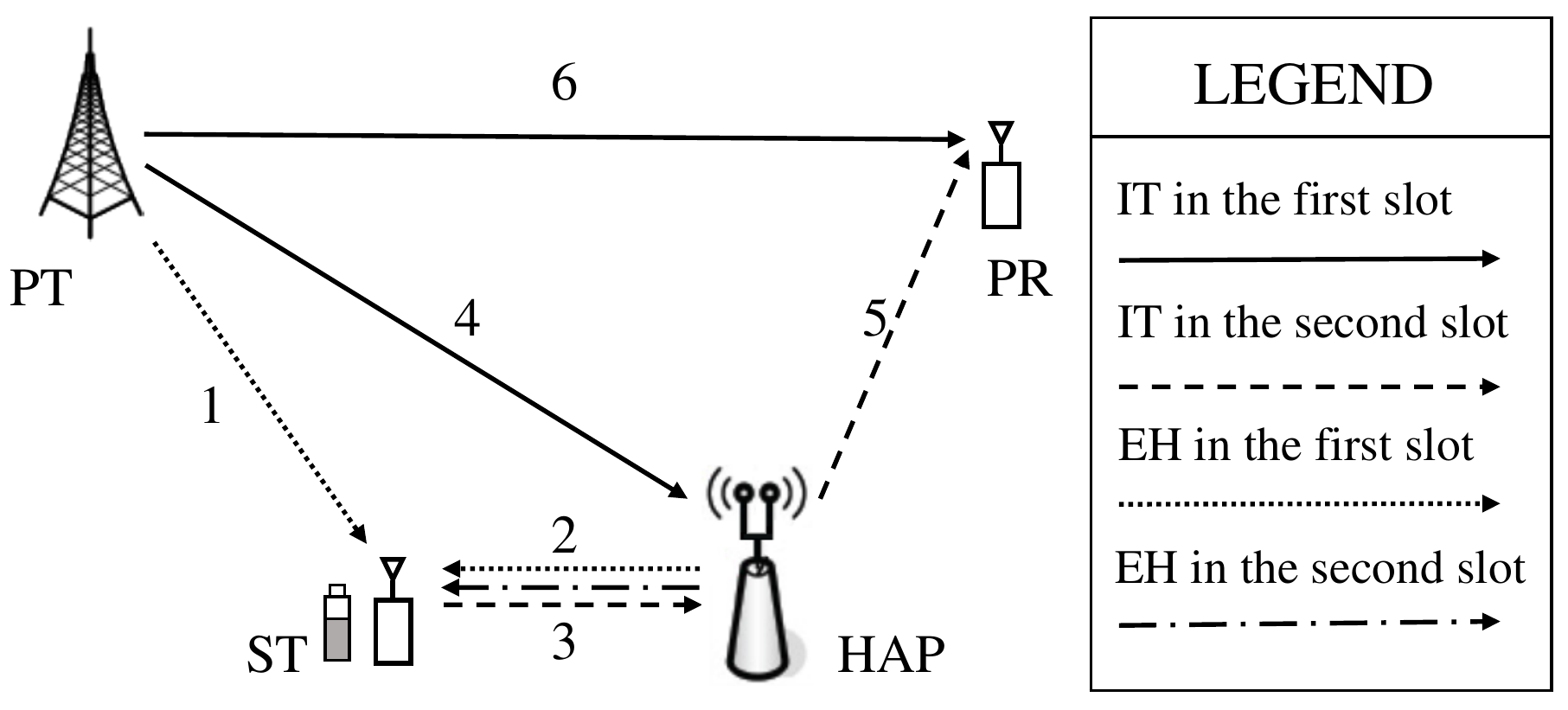}
		\caption{System model of the considered FD cooperative cognitive radio network with wireless EH. We assume Nakagami-$m$ fading with parameters $m_i$ and $\Omega_i$, where $i \in \{1;2;3;4;5;6\}$ denotes links as marked in Fig. 1.}
		\label{system model}
	\end{figure}

	We consider a FD cooperative cognitive radio network with wireless EH consisting of a PT-PR pair, a ST and a FD secondary HAP, as shown in Fig. \ref{system model}. In line with \cite{7342976,6847192,7434602}, we assume that the PT, PR, and HAP are connected to external power supplies, while the ST has no embedded power supply. Instead, the ST is equipped with a wireless EH unit and a finite-capacity battery such that it can harvest and accumulate energy from the RF signals transmitted by the PT and HAP. This system fits well with future Internet-of-Things applications where the HAP processes information gathered from low power STs.
	
	In this paper, we propose a new CSS protocol for the considered network. To elaborate, we use $T$ to denote the duration of each transmission block, which is further divided into two equal time slots with length $T/2$. For the cooperative information transmission of the primary network, the PT broadcasts information to the PR and HAP in the first time slot. In the second time slot, the HAP relays the PT information to the PR, which combines the signals received from the PT and HAP. For the secondary network, the ST harvests energy from the PT and HAP in the first time slot. Thanks to its FD capability, the HAP can simultaneously receive information from the PT and transmit a predefined energy-bearing RF signal to charge the ST in the first time slot. In the second time slot, if the residual energy in the ST battery is above a predefined threshold $E_t$, the ST will consume $E_t$ amount of accumulated energy to transmit its own information to the HAP. The FD HAP receives the signals from the ST and relays the PT's information to PR at the same time. If the energy in the battery is below the threshold, the ST will continue to harvest energy from the information relaying signals emitted by the HAP. For simplicity, we adopt the fixed decode-and-forward (DF) relaying protocol at the HAP.
	
	Hereafter, we use $h_{\mathcal{S}\mathcal{T}},\mathcal{S},\mathcal{T} \in \left\lbrace {a,b,c,d}\right\rbrace $ to represent the complex channel coefficient of the link between $\mathcal{S}$ and $\mathcal{T}$, where ${a,b,c,d}$ denote the nodes PT, PR, ST, and HAP, respectively, and $\gamma_{\mathcal{ST}}$ denotes the received signal-to-interference-plus-noise ratio (SINR) from $\mathcal{S}$ to $\mathcal{T}$. We note that the line-of-sight (LoS) path is very likely to exist in wireless EH links, thus we use Nakagami-$m$ fading model to characterize the channel fading for all links, which is mathematically tractable and can well approximate the Rician fading model for LoS paths. Furthermore, all links in the system are assumed to remain unchanged over each transmission block and varies from one block to another independently following a Nakagami-$m$ fading distribution with shape parameter $m_i$, and square value of power gains $\Omega_{i}$, where $i \in \{1,2,3,4,5,6\}$ is used to denote different links as marked in Fig. \ref{system model}.
	
	We now further detail the proposed CSS protocol and mathematically describe the harvested energy at the ST, and the received SINRs at the PR and HAP. Without loss of generality, we hereafter consider a normalized transmission interval (i.e., $T$ = 1). For EH in the first time slot, the amount of energy harvested by the ST is given by \cite{6623062}
	\begin{equation}
	\label{equ_1}
	E_{H} = \frac{1}{2} \eta \left( P_{{a}} \left| h_{{ac}} \right|^2 + P_{{d}} \left| h_{{cd}} \right|^2 \right),
	\end{equation}
	where $\eta$ is the energy conversion efficiency at the ST, and $P_{\mathcal{S}},\mathcal{S} \in \{a,d\}$, denotes the transmit power of node $\mathcal{S}$. Notice that $E_{H}$ is multiplied by a factor of 1/2 as the ST harvests energy from only half of the transmission block. The noise power is not considered here as it is normally below the sensitivity of wireless EH units.
	
	For IT in the first time slot, the received SINRs at the PR and HAP are given by
	\begin{equation}
	\gamma_{{ab}} = P_{{a}} \left| h_{{ab}} \right|^2/N_0,
	\end{equation}
	and
	\begin{equation}
	\gamma_{{ad}} = P_{{a}} \left| h_{{ad}} \right|^2/(N_0 + P_{{d}} H_{{dd}}),
	\end{equation}
	respectively, where $N_0$ represents the noise power, and $H_{dd}$ represents the loop interference at the HAP that remains after imperfect loop interference elimination \cite{5961159}. We assume $H_{dd}$ is a constant value in this paper. This is motivated by the fact that the energy-bearing signal can be effectively reduced by advanced analog and digital self-interference cancellation, thus $H_{{dd}}$ can be very small and the randomness can be suppressed dramatically. We assume that the EH signals from the HAP are predefined and can be cancelled completely at the PR, thus the information transmission from the HAP to the PR will not be interfered by the energy-bearing signals emitted by the HAP.
	
	In the second time slot, the HAP relays the received information to the PR using fixed DF protocol, while the ST chooses to operate in one of the two following modes depending on the amount of energy in its battery.
	
	\subsection{Mode I: Information Transmission at the ST}
	In this mode, the accumulated energy available in the ST battery reaches the predefined threshold $E_t$, thus the ST transmits information to the HAP by consuming $E_t$ amount of energy from its battery. The received SINR at the HAP can be expressed as
	\begin{equation}
	\label{equ_4}
	\gamma_{{cd}} = \frac{2E_t |h_{{cd}}|^2}{N_0 + P_{{d}} H_{{dd}}},
	\end{equation}
	where the factor 2 comes from the fact that the $E_t$ amount of energy is consumed within half a time block. At the same time, the PR receives information from the HAP, which is interfered by the signal from the ST. The received SINR at the PR can thus be given by
	\begin{equation}
	\gamma_{db,I} = \frac{P_d |h_{{db}}|^2}{N_0 + 2E_t |h_{{cb}}|^2}.
	\end{equation}
	In mode I, the ST only harvests energy during the first time slot, thus the total amount of harvested energy at the ST can be given by $E_I = E_H$.
	
	\subsection{Mode II: Energy Harvesting at the ST}
	In this mode, the accumulated energy available in the ST battery is below $E_t$ after the first slot, thus the ST will continue to harvest energy from the RF signal emitted by the HAP for information relaying. In this mode, the total amount of energy harvested in the two time slots can be expressed as
	\begin{equation}
	\label{equ_2}
	E_{II} = E_H + \frac{\eta}{2} P_{{d}} \left| h_{{cd}} \right|^2 = \frac{\eta}{2} \left( P_{{a}} \left| h_{{ac}} \right|^2 + 2 P_{{d}} \left| h_{{cd}} \right|^2 \right).
	\end{equation} 
	Moreover, the PR will not suffer from interference in the second time slot and the SINR at the PR is thus given by
	\begin{equation}
	\gamma_{{db,II}} = P_d |h_{{db}}|^2/N_0.
	\end{equation}
	
	For both mode I and II, we consider that the PR will implement maximum ratio combining (MRC) to combine the signals received in the first and second time slots. As such, the total SINR at the PR can be written~as
	\begin{equation}
	\label{equ_8}
	\gamma_{{b,i}} = \gamma_{{ab}} + \min\left( \gamma_{ad},\gamma_{{db,i}} \right),
	\end{equation}
	where $i \in \{I,II\}$.
	
	\section{Performance Analysis}
	In this section, we will analyze the achievable throughput of the proposed CSS protocol over Nakagami-$m$ fading channels. To this end, we follow \cite{7417666} to model the dynamic behaviors of the ST battery as a finite-state Markov chain.
	\subsection{Markov Chain Description of ST Battery}
	We consider that the ST is equipped with a $L+1$ discrete-level battery with a finite capacity $C$. The $i$th energy level is defined as $\varepsilon_i = {iC}/{L}$, $i\in \{0,1,...,L\}$. Note that when the value of $L$ is large enough, the discrete-level battery can tightly approximate the actual continuous behaviors of the ST battery \cite{4524272}.  With the adopted discrete battery model, the discretized amount of harvested energy at the ST in mode I is given by
	\begin{equation}
	\varepsilon_{I} \triangleq \varepsilon_i,  \text{where }  i = \arg \max_{j \in 0,1,...,L} \{\varepsilon_j:\varepsilon_j < E_{I}\},
	\end{equation}
	and the harvested energy at the ST in mode II is given by
	\begin{equation}
	\varepsilon_{II} \triangleq \varepsilon_i,  \text{where }  i = \arg \max_{j \in 0,1,...,L} \{\varepsilon_j:\varepsilon_j < E_{II}\}.
	\end{equation}
	
	Let $t$ denote the energy level corresponding to the discretized energy threshold, which can be expressed as
	\begin{equation}
	t \triangleq i,  \text{where }  i = \arg \min_{j \in 1,...,L} \{\varepsilon_j:\varepsilon_j \geqslant E_{t}\}.
	\end{equation}

	The residual energy at the beginning of the $\left(n+1\right)$-th transmission block can thus be expressed as
	\begin{equation}
	\varepsilon \left[n + 1\right] = 
	\begin{cases}
	\min \left\{\varepsilon \left[n\right] + \varepsilon_I, C\right\} - \varepsilon_{t} , &\mbox{if $\varepsilon \left[n\right] + \varepsilon_{I} \geqslant \varepsilon_{t} $}; \\
	\min \left\{\varepsilon \left[n\right] + \varepsilon_{II}, C\right\}, & \mbox{if $ \varepsilon \left[n\right] + \varepsilon_{I} < \varepsilon_{t} $},
	\end{cases}
	\end{equation}
	where $\varepsilon [n] $ represents the residual energy in the ST battery at the beginning of the $n$-th transmission block.
	
	Inspired by \cite{7417666}, we adopt the Markov chain (MC) to model the dynamic behavior of the ST battery. We use $S_i$ to denote the state of the ST battery being $\varepsilon_i$ at the beginning of each transmission block, and $V_{i,j}$ to denote the transition probability from state $S_i$ to $S_j$. The state transitions of the MC are summarized in the following cases.
	
	\subsubsection{Case 1: \textit{The battery remains uncharged ($S_i$ to $S_i$ with $0 \leqslant i \textless L-t$).}}
	 In this case, the incremental amount of energy (i.e., $\varepsilon_{I}-\varepsilon_t$ in Mode I or $\varepsilon_{II}$ in Mode II) at the ST battery should be less than $\varepsilon_1$ so that it can be discretized to zero. Thus the transition probability can be characterized as
	\begin{equation}
	\small
	\begin{split}
	V_{i,i} = & \Pr \bigg \{ \left[ (\varepsilon_{I} = \varepsilon_t ) \cap \left( \varepsilon_{t}  \leqslant \varepsilon_i + \varepsilon_{I} < \varepsilon_{L} \right) \right] \\
	& \cup \left[ \left( \varepsilon_{II} = 0 \right) \cap \left(0 \leqslant \varepsilon_i + \varepsilon_I < \varepsilon_{t} \right)  \right] \bigg \} \\
	= &
	\begin{cases}
	F_{{II}}\left(\varepsilon_1\right) + F_{{I}}\left(\varepsilon_{t+1}\right)- F_{{I}}\left(\varepsilon_{t}\right),  & \mbox{if $0 \leqslant i \leqslant t-1$}; \\
	F_{{I}}\left(\varepsilon_{t+1}\right)- F_{{I}}\left(\varepsilon_{t}\right), & \mbox{if $t \leqslant i < L-t$},
	\end{cases}
	\end{split}
	\end{equation}
	where $F_{{I}}(\cdot)$ and $F_{{II}}(\cdot)$ denote the cumulative distribution functions (CDF) of $E_{I}$ and $E_{II}$, respectively. Based on the results presented in \cite{1673666}, we can derive closed-form expressions for $F_I(\cdot)$ and $F_{II}(\cdot)$ expressed as
	\begin{equation}
	\label{F_I}
	F_{I}(x) =  \frac{\eta N_0}{2} \sum_{\mu = 1}^{2} \sum_{\nu = 1}^{m_{\mu}} \frac{A(\mu,\nu) }{\Gamma(m_{\mu})} \gamma(m_{\mu}, \frac{2 x \beta_{\mu}}{\eta N_0}),
	\end{equation}
	and
	\begin{equation}
	\label{F_II}
	F_{II}(x) = \frac{\eta N_0}{2}\sum_{\mu = 1}^{2} \sum_{\nu = 1}^{m_{\mu}} \frac{A(\mu,\nu) \gamma(m_{\mu},2^{-U(\mu-2)} \beta_{\mu} x)}{\Gamma(m_{\mu})},
	\end{equation}
	where we consider all $m_i$, $i \in \{1,2,3,4,5,6\}$, as integers for simplicity, $U(x)$ is the unit step function defined as $U(x)=1$ for $x \geqslant 0$, and zero otherwise, and the weight $A(\cdot,\cdot)$ is given in \eqref{Equ_1} on top of this page.
	\begin{figure*}[!t]
		\normalsize
		\begin{equation}
		\label{Equ_1}
		A(\mu,\nu) = (-1)^{m_1+m_2-m_{\mu}}\frac{\beta_1^{m_1} \beta_2^{m_2} \Gamma(m_1 + m_2 - \nu)(\beta_{\mu} - \beta_{1+U(1-\mu)})^{\nu - m_1 - m_2}}{\beta_{\mu}^{\nu}\Gamma(m_{1+U(1-\mu)}) \Gamma(m_{\mu} - \nu + 1)}.
		\end{equation}
		\hrulefill
	\end{figure*}
	In \eqref{F_I} and \eqref{F_II}, $\gamma(\cdot,\cdot)$ is the lower incomplete gamma function, and $\beta_{i} = m_{i}/\Omega_{i}, i \in \{1,2\}$. In our proposed model, $\Omega_1$ and $\Omega_2$ are defined as
	\begin{equation}
%	\Omega_1 = \frac{P_a d_{ac}^{-\alpha}}{N_0},\\
%	\Omega_2 = \frac{P_d d_{cd}^{-\alpha}}{N_0},
	\Omega_1 = P_a /(d_{ac}^{\alpha} N_0),\\
	\Omega_2 = P_d /(d_{cd}^{\alpha} N_0),
	\end{equation}
	where $\alpha$ is the path loss exponent and $d_{\mathcal{ST}}$ is the distance between nodes $\mathcal{S}$ and $\mathcal{T}$.
	
	\subsubsection{Case 2: \textit{The battery is partially charged ($S_i$ to $S_j$ with $0 \leqslant i < j \leqslant L$).}} Similar with the previous case, the harvested energy should be larger than $\varepsilon_{j-i+t}$ in Mode I or between $\varepsilon_{j-i}$ and $\varepsilon_{j-i+1}$ in Mode II. Thus, this transition probability can be characterized as \eqref{Equ_2}
	\begin{figure*}[!t]
		\normalsize
		\begin{equation}
		\label{Equ_2}
		\begin{split}
		V_{i,j} = &
		\mathrm{Pr}\bigg \{ \left[ \left(\varepsilon_{II} = \varepsilon_{j-i} \right) \cap \left( 0 \leqslant \varepsilon_i + \varepsilon_{I} < \varepsilon_{t} \right) \right] \cup [ \left(\varepsilon _{I} = \varepsilon_{j-i+t} \right) \cap \left( \varepsilon_{t} \leqslant \varepsilon_i + \varepsilon_{I} < \varepsilon_{L-1} \right) ] \bigg \} \\
		= &
		\begin{cases}
		F_I(\varepsilon_{t+j-i+1}) - F_I(\varepsilon_{t+j-i}) + F_{II}(\varepsilon_{j-i+1}) - F_{II}(\varepsilon_{j-i}), & \mbox{if $0 \leqslant i \leqslant t-2$ and $i+1 \leqslant j \leqslant t-1$}; \\
		F_I(\varepsilon_{t+j-i+1}) - F_I(\varepsilon_{t+j-i}) + \mathcal{F}(i,j,\beta_1,\beta_2,m_1,m_2,t,\varepsilon_1), & \mbox{if $0 \leqslant i \leqslant t-1$ and $t \leqslant j \leqslant L-t-1$};\\
		F_I(\varepsilon_{t+j-i+1}) - F_I(\varepsilon_{t+j-i}) , & \mbox{if $t \leqslant i < j \leqslant L-t-1$};\\
		1 - F_I(\varepsilon_{t+L-i}) , & \mbox{if $0 \leqslant i \leqslant L-t$ and $j = L-t$},
		\end{cases}
		\end{split}
		\end{equation}
		\hrulefill
	\end{figure*}
	on top of this page, where $\mathcal{F}(i,j,\beta_1,\beta_2,m_1,m_2,t,\varepsilon_1)$ is given by
	\begin{equation}
	\label{Equ_3}
	\begin{split}
	& \mathcal{F} (i,j,\beta_1,\beta_2,m_1,m_2,t,\varepsilon_1) \\
	= & \Pr \bigg \{ \left(\varepsilon_i + \varepsilon_{I} < \varepsilon_t\right) \cap \left(\varepsilon_j = \varepsilon_i + \varepsilon_{II} \geqslant \varepsilon_t\right) \bigg \} \\
	= & \sum_{\mu=j-t+1}^{\lfloor (j-i)/2 \rfloor} P_{1}(j-i-2\mu) P_{2}(\mu),
	\end{split}
	\end{equation}
	In \eqref{Equ_3}, $\lfloor x \rfloor$ denotes the largest integer that is smaller than $x$, and $P_{i}(x), i \in \{1,2\}$, denotes the probability that the ST harvests $x\varepsilon_1$ amount of energy through link $i$ in a half transmission block, which can be derived as
	\begin{equation}
	P_{i}(x) = \frac{\gamma(m_i, \frac{2 \beta_i \varepsilon_{x+1}}{\eta N_0})}{\Gamma(m_i)} - \frac{\gamma(m_i,\frac{2 \beta_i \varepsilon_x}{\eta N_0})}{\Gamma(m_i)}.
	\end{equation}
	
	\subsubsection{Case 3: \textit{The battery is fully charged ($S_i$ to $S_L$ with $0 \leqslant i < 2t-L$ and $2t > L$).}} This case happens only when ST operates in Mode II, as the ST battery can never be fully charged in Mode I. We can further deduce that the residual energy in the battery should be less than the energy threshold after the first slot but reaches the battery capacity after the second slot. Therefore, this transition probability can be expressed~as
	\begin{equation}
	\begin{split}
	V_{i,L} = &
	\mathrm{Pr}\bigg \{ \left[ \left(\varepsilon_{i} + \varepsilon_{I} < \varepsilon_{t} \right) \cap \left( \varepsilon_i + \varepsilon_{II} \geqslant \varepsilon_{L} \right) \right]  \bigg\} \\
	= & \sum_{\mu=L-t}^{\lfloor\frac{L-i}{2} \rfloor} \sum_{\nu=L-i-2\mu}^{t-i-\mu-1} P_1(\mu) P_2(\nu), 0 \leqslant i < 2t-L.
	\end{split}
	\end{equation}
	
	\subsubsection{Case 4: \textit{The battery is discharged ($S_j$ to $S_i$ with $1 \leqslant j \leqslant L-t$ and $ \max\left(0,j-t \right)  \leqslant i < j$).}} It is obvious that the energy in the ST battery will decrease only when the residual energy of the ST reaches $E_t$ after the first slot but the total harvested energy is smaller than the amount of consumed energy (i.e., the energy threshold) in Mode I. Thus, the transition probability of this case is characterized~as
	\begin{equation}
	\begin{split}
	V_{j,i} = &
	\mathrm{Pr}\bigg \{ \left[ \left(\varepsilon _{I} = \varepsilon_t - \varepsilon_{j-i} \right) \cap \left( \varepsilon_t \leqslant \varepsilon_j \leqslant \varepsilon_{L} \right) \right] \bigg\} \\
	= & F_I\left(\varepsilon_{t-j+i+1}\right) - F_I\left(\varepsilon_{t-j+i}\right), \\
	& \mbox{if $1 \leqslant j \leqslant L-t$ and $\max\left(0,j-t \right)  \leqslant i < j$}.
	\end{split}
	\end{equation}
	
	Let $\mathbf {V}=[V_{i,j}]_{\left(L+1 \right) \times \left( L+1\right)}$ denote the transition matrix of the formulated MC, which can easily be verified to be irreducible and row stochastic \cite{6177989}. Therefore, there exists a unique solution $\bm{\pi}$ that satisfies the following equation 
	\begin{equation}
	\bm{\pi} = (\pi_{0},\pi_{1},...,\pi_{L})^\mathrm{T} = \mathbf{V}^\mathrm{T} \bm{\pi},
	\end{equation}
	where $\pi_i$, $i \in \{0,1,...,L\}$, is the $i$-th component of $\bm{\pi}$ representing the stationary distribution of the $i$-th energy level at the ST at the beginning of each transmission block. The battery stationary can be calculated~as~\cite{6177989}
	\begin{equation}
	\bm{\pi} = (\mathbf{V}^\mathrm{T}  - \mathbf{I} + \mathbf{B})^{-1} \mathbf{b},
	\end{equation}
	where $\mathbf{V}^\mathrm{T}$ denotes the transpose matrix of $\mathbf{V}$, $\mathbf{I}$ is the identity matrix, $\mathbf{B}_{i,j} = 1, \forall i,j$, and $\mathbf{b} = (1,1,...,1)^\mathrm{T}$.

	\subsection{Achievable Throughput Analysis}
	Based on the steady state distribution of the ST battery derived in the previous subsection, we now proceed to analyze the achievable throughput of the proposed protocol. From \eqref{equ_4}, the achievable throughput at the HAP can be expressed as 
	\begin{equation}
	R_{{d}} = (1 - P_{od}) R_0 / 2,
	\end{equation}
	where $R_0 = log_2(1+\gamma_0)$, and $\gamma_0$ refers to the SINR threshold. The $P_{od}$ denotes the probability of the event that the energy in ST battery is smaller than $E_t$ or the transmission of ST suffers from outage, which can be expressed as
	\begin{equation}
	\begin{split}
	P_{od}  = & 1 - \mathrm{Pr} \left\{ \gamma_{{cd}} \geqslant \gamma_0 \ \& \ \varepsilon_{II} \geqslant \varepsilon_{t-i} \right\} \\
	= & \sum_{\nu=0}^{s} P_2(\nu) + \sum_{i=0}^{t-s} \sum_{\mu=0}^{t-i} \sum_{\nu=s}^{t-\mu-i} \pi_i P_1(\mu) P_2(\nu),
	\end{split}
	\end{equation}
	where $s$ denotes the minimum battery level of the energy harvested from HAP signals. Through combining (\ref{equ_1}), (\ref{equ_4}), and (\ref{equ_2}), $s$ can be given by
	\begin{equation}
	s = \lceil \frac{\eta P_d \gamma_0 L}{4 C E_t} (N_0 + P_d H_{dd}) \rceil,
	\end{equation}
	where $\lceil x \rceil$ denotes the smallest integer that is larger than x.
	
	Based on \eqref{equ_8}, the capacity of the PR is expressed as \cite{6827192}
	\begin{equation}
	\label{equ_36}
	R_b = \frac{R_0}{2} \left[ (1 - P_{ob}) (1 - F_{\gamma_{{b,I}}}(x)) +  P_{ob} (1 - F_{\gamma_{{b,II}}}(x)) \right],
	\end{equation}
	where $F_{\gamma_{b,I}}(\cdot)$ and $F_{\gamma_{b,II}}(\cdot)$ are the CDF of $\gamma_{{b,I}}$ and $\gamma_{{b,II}}$, respectively, and $P_{ob}$ denotes the probability of the event that the residual energy of the ST battery is smaller than the energy threshold in the middle of each transmission~block.
	
	We find that the exact CDF of $\gamma_{{b,I}}$ is difficult to derive. Fortunately, we realized that the interference in $\gamma_{{b,I}}$ could be small since its transmit power comes from EH, and ST and PR can be far away from each other. For tractability, we omit the interference term in $\gamma_{b,I}$ such that $\gamma_{b,I} \approx \gamma_{b,II}$. We then derive the CDF of $\gamma_{b,I}$ given in \eqref{Equ_5} on top of this page
	\begin{figure*}[!t]
		\normalsize
		\begin{equation}
		\label{Equ_5}
		F_{\gamma_{b,I}}(x) = 1 - e^{-\beta_6 x} - \sum_{a = 0}^{m_4-1} \sum_{b = 0}^{m_5 - 1}\frac{\beta_4^a \beta_5^b \beta_6^{m_6}}{a! b! \Gamma(m_6)} B(a + b + 1,m_6) e^{-(\beta_4 + \beta_5)x} x^{a + b + m_6} \ _1 F_1(m_6; a+b+1;x(\beta_4+\beta_5-\beta_6)),
		\end{equation}
		\hrulefill
	\end{figure*}
	where 
	$\beta_i = m_i/\Omega_i, i \in \{4,5,6\}$, which denotes the different links between nodes, and the values of $\Omega_4$, $\Omega_5$, and $\Omega_6$ are defined~as
	\begin{equation}
	\Omega_4 = \frac{P_a d_{ad}^{-\alpha}}{N_0 + P_d H_{dd}}, \Omega_5 = \frac{P_d d_{db}^{-\alpha}}{N_0}, \Omega_6 = \frac{P_a d_{ab}^{-\alpha}}{N_0}.
	\end{equation}
	
	\section{Numerical Results}	
	In this section, we provide simulation results to verify the correctness of the derived analytical expressions and illustrate the impacts of key parameters on the achievable throughput of the proposed CSS protocol. For simplicity, we consider a linear network topology, where the nodes are located along a straight line according to the following order: PT, ST, HAP, PR. Therefore, we have $d_{ad} = d_{ac} + d_{cd}$ and $d_{ab} = d_{ad} + d_{bd}$. In all simulations, we consider that the transmit powers of the PT and HAP follows a specific relationship given by $P_a = k P_d$, where $k$ is a constant. We set the distance between the PT and PR as $d_{ab} = 20$, path-loss factor $\alpha = 3$, energy conversion efficiency $\eta = 0.5$, noise power $N_0 = 10^{-5}$, ST battery capacity $C = 5$, and Nakagami-$m$ fading parameter $m_i = 3$, for $i \in \{1,2,3\}$, and $m_i = 1$, for $i \in \{4,5,6\}$.
	
	\begin{figure}[h]
		\centering
		\includegraphics[width=0.48\textwidth]{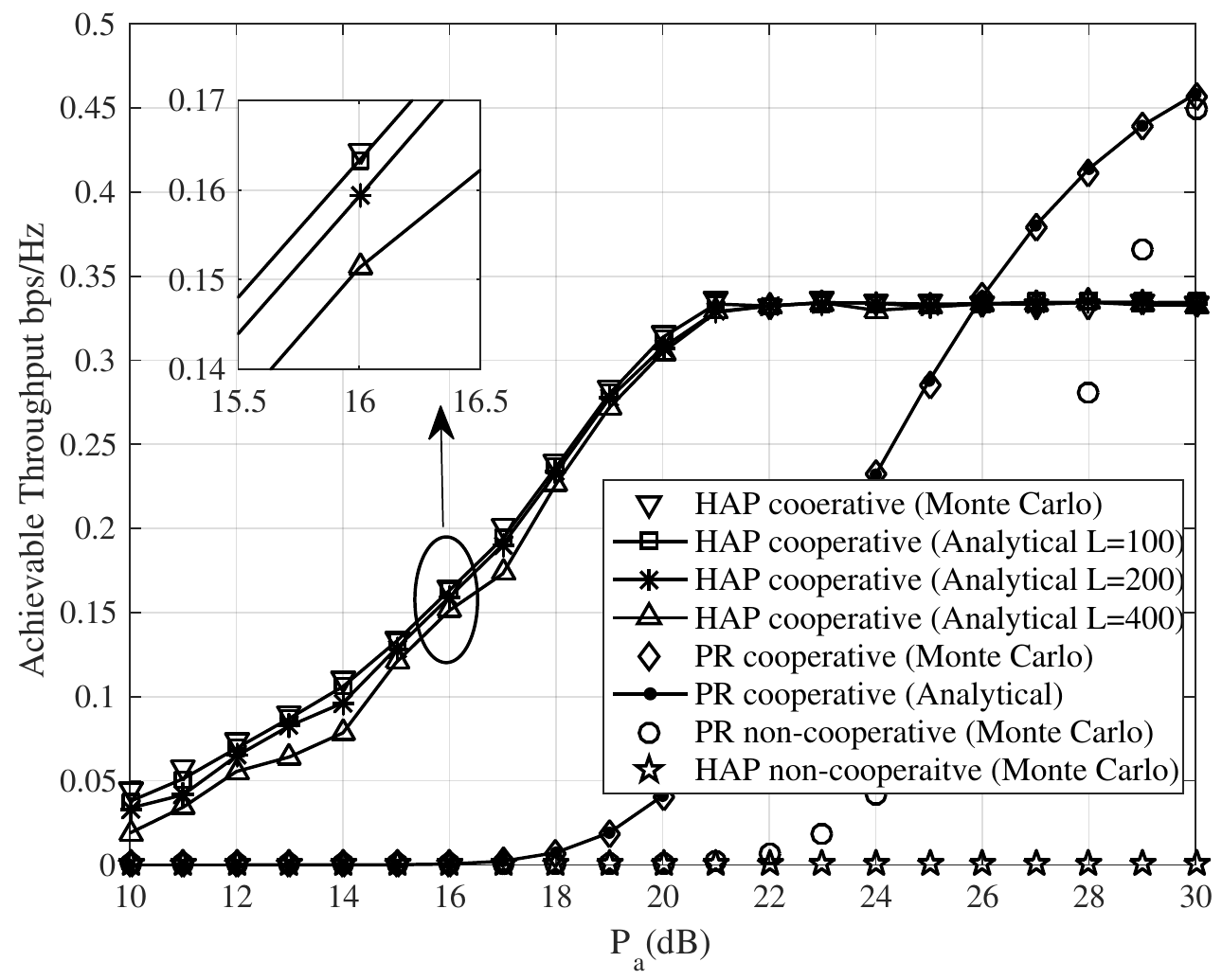}
		\caption{The achievable throughput of the PR and HAP versus the transmit power at the PT with $E_t = 2$, $d_{{ad}}=6$, $d_{{ac}}=d_{{cd}}=3$, and $P_{a} = P_{d}$.}
		\label{fig_1}
	\end{figure}
	
	Fig. \ref{fig_1} compares the achievable throughput of the PR and HAP for the considered cooperative network and its non-cooperative counterpart. We can see from this figure that the analytical results and the corresponding Monte Carlo simulations of the achievable throughput at the PR match well with each other. Moreover, similar phenomenon can be observed for the average throughput at the HAP for a wide range of value of $P_a$. It is worth pointing out that when $P_a$ is relatively small, there exist small gaps between the analytical and simulation curves for the average throughput of the HAP. This is because the harvested energy in the Monte Carlo simulations is continuous, whereas it is discretized in our analytical results. When $P_a$ is small, the amount of energy harvested at each time is less than a single discrete battery level and is thus treated as zero in our analytical results. In contrast, the amount of energy at the ST battery indeed increases in the Monte Carlo simulations.
	
	Fig. \ref{fig_1} also shows that the average throughput of the PR increases monotonically with the PT transmit power. We see that the average throughput of the PR in the proposed cooperative protocol significantly outperforms the non-cooperative case. As we assume the ST is not permitted to transmit information in the non-cooperative network, the average throughput at the HAP in the proposed cooperative framework also outperforms the non-cooperative counterpart. We can also observe that the achievable throughput at the HAP achieves the maximum value when $P_a$ is greater than 22$dBm$. This is because when the PT and HAP are in the low transmit power regime, the probability of the two events where the ST battery level is smaller than $E_t$ and the SINR at HAP is smaller than $\gamma_0$, both decrease with increasing $P_a$ and $P_d$. However, the achievable throughput of the ST will remain stable when $P_a$ and $P_d$ are large enough and the ST can transmit information to the HAP in each transmission~block.
	
	\begin{figure}[h]
		\centering
		\includegraphics[width=0.48\textwidth]{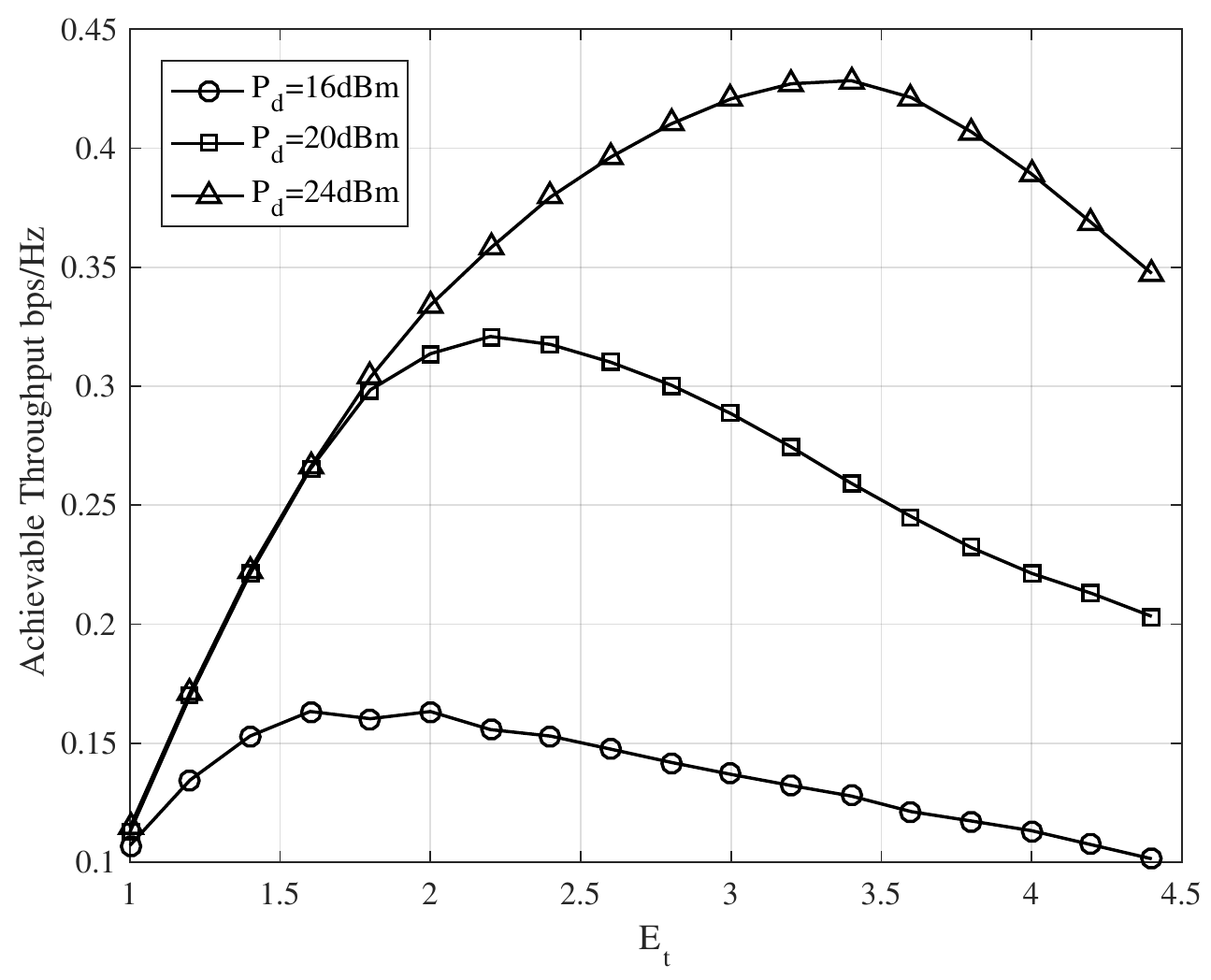}
		\caption{The achievable throughput of the HAP versus the energy thresholds with different PT transmit powers, where $d_{{ad}}=6$, $d_{{ac}}=3$, $d_{{cd}}=3$, $P_{{a}} = 
		P_{d}$, and $L=400$.}
		\label{fig_3}
	\end{figure}
	
	Fig. \ref{fig_3} shows the achievable throughput of the HAP versus the energy threshold $E_t$ with different values of $P_d$. We can see that in all simulated cases, there exists an optimal $E_t$ that maximizes the achievable throughput of the HAP. This is because the $E_t$ has contradictory influences on the HAP's throughput. A larger $E_t$ increases the HAP's instantaneous achievable throughput but reduces the transmission probability of the ST. We also observe from the figure that the optimal value of $E_t$ shifts to the right as the $P_d$ increases. This is because the ST can harvest more energy on average and thus a higher $E_t$ should be adopted to increase the throughput at the HAP.
	
	\section{Conclusions}
	In this paper, we developed a novel cooperative spectrum sharing framework for a full-duplex cooperative cognitive radio network with wireless energy harvesting (EH). The full-duplex hybrid access point (HAP) can simultaneously receive signals from the primary transmitter (PT) and emit signals to charge the secondary transmitter (ST) in the first time slot, and can forward the PT's signals to the primary receiver (PR) and receive information from the ST in the second time slot. The energy accumulation process of the EH ST was modeled as a finite-state Markov chain and analytical expressions were derived for the achievable throughput of both primary and secondary links. Monte Carlo simulations validated our theoretical analysis. Numerical results demonstrated that the achievable throughput of the proposed cooperative framework outperforms that of its non-cooperative counterpart.

	\bibliography{document}
	\bibliographystyle{IEEEtran}
	
\end{document}